# Revealing Critical Characteristics of Mobility Patterns in New York City during the Onset of COVID-19 Pandemic

**Akhil Anil Rajput[1*], Qingchun Li[2], Xinyu Gao[3], Ali Mostafavi[4]**

[1]Ph.D. Student, UrbanResilience.AI Lab, Zachry Department of Civil and Environmental Engineering, Texas A&M University, College Station, USA 77843-3136; email: akhil.rajput@tamu.edu

[2]Ph.D. Student, UrbanResilience.AI Lab, Zachry Department of Civil and Environmental Engineering, Texas A&M University, College Station, USA 77843-3136; email: qingchunlea@tamu.edu

[3]Ph.D. Student, UrbanResilience.AI Lab, Zachry Department of Civil and Environmental Engineering, Texas A&M University, College Station, USA 77843-3136; email: xy.gao@tamu.edu

[4]Assistant Professor, UrbanResilience.AI Lab, Zachry Department of Civil and Environmental Engineering, Texas A&M University, College Station, USA 77843-3136; email: amostafavi@tamu.edu

**\*** Corresponding Author. Email address: akhil.rajput@tamu.edu (Mr. Akhil Anil Rajput)

## Abstract

New York has become one of the worst-affected COVID-19 hotspots and a pandemic epicenter due to the ongoing crisis. This paper identifies the impact of the pandemic and the effectiveness of government policies on human mobility by analyzing multiple datasets available at both macro and micro levels for the New York City. Using data sources related to population density, aggregated population mobility, public rail transit use, vehicle use, hotspot and non-hotspot movement patterns, and human activity agglomeration, we analyzed the inter-borough and intra-borough moment for New York City by aggregating the data at the borough level. We also assessed the internodal population movement amongst hotspot and non-hotspot points of interest for the month of March and April 2020. Results indicate a drop of about 80% in people's mobility in the city, beginning in mid-March. The movement to and from Manhattan showed the most disruption for both public transit and road traffic. The city saw its first case on March 1, 2020, but disruptions in mobility can be seen only after the second week of March when the shelter in place orders was put in effect. Owing to people working from home and adhering to stay-at-home orders, Manhattan saw the largest disruption to both inter- and intra-borough movement. But the risk of spread of infection in Manhattan turned out to be high because of higher hotspot-linked movements. The stay-at-home restrictions also led to an increased population density in Brooklyn and Queens as people were not commuting to Manhattan. Insights obtained from this study would help policymakers better understand human behavior and their response to the news and governmental policies.

# 1. INTRODUCTION

With more than 90 million confirmed cases and 1.9 million deaths worldwide at the time of this writing, the COVID-19 global pandemic has caused unprecedented social, economic, and environmental impacts (WHO Coronavirus Disease (COVID-19) Dashboard, 2021; Bashir et al., 2020). New York City, which became a major hotspot in the United States during the early stages of the pandemic, had more than 500,000 confirmed cases as of January 15, 2021. New York City observed its first peak during the first week of April. Public policy and implementation of control measures are essential to support social distancing, which might help slow the spread of COVID-19 (Lasry et al., 2020; Kraemer et al., 2020). Huang et al. (2020a) have shown that mobility changes correspond well with the declaration of mitigation measures, implying effectiveness. Additionally, a study done in the U.K. analyzed the impact of government control measures on human mobility reduction and identified a relationship between human mobility trends and COVID-19 cases (Hadjidemetriou et al., 2020). They found out that reduction in mobility led to lower COVID-19 related casualties.

This study assesses the impact of the COVID-19 pandemic and government policies on human mobility for four major boroughs of New York City. The city was an early epicenter of the COVID-19 pandemic in the United States (Thompson et al., 2020). New York City is also one of the most densely populated cities in the United States, has a well-established intracity public rail, and diversity in transport options. Studies have shown that human mobility correlates directly with the number of positive cases with some time lag (Glaeser et al., 2020; Carteni et al., 2020; Badr et al., 2020; Xiong et al., 2020) and disease reproduction number (Linka et al., 2020). Iacus et al. (2020) found that mobility alone can explain more than 90% of the initial spread of the COVID-19 virus. Furthermore, Gatto et al. (2020) have found that policies restricting population movement in Italy and resulting lessened mobility reduce virus transmission by 45%. Yabe et al. (2020) found that even non-compulsory measures in Tokyo resulted in a 50% drop in mobility and led to 70% fewer social contacts. Therefore, understanding the effectiveness of government policies on reducing human mobility will help shape these policies and better control any future outbreaks.

Almagro et al. (2020) found that crowded spaces play a more critical role than population density in spreading the infection. Gao et al. (2020) used Venables distance as a means to understand the agglomeration of human activities at the county level for 194 US counties. Li et al. (2020) classified points of interest (POIs) as hotspots or non-hotspots and assessed the mobility patterns among them for US cities. They found that while visits to hotspots decreased in some cities, some did not show a considerable drop; however, these studies did not focus on a granular level. Our study was focused on finding these measures at the borough level for New York City to understand whether boroughs show any disparity in movement patterns. We aggregated the POIs at the borough level for New York City to assess the inter-borough and intra-borough hotspot and non-hotspot mobility trends. This knowledge may enable policymakers to better manage the control measures for different boroughs.

Researchers have also analyzed the effect of mobility on greenhouse gas emissions by the aviation and transport sectors (Abu-Rayesh and Dincer, 2020). For instance, Jiang et al. (2020) identified the impact of human behavior and mobility on Singapore's environment. They also reported a 30% reduction in mobility leads to about 44% to 55% reduction in air emissions related to transportation.

Zhou et al. (2020) used mobile phone data to build an exposure model for Shenzhen, China, and found that the reduction in mobility helped flatten the peak number of cases and lead to a delayed peak.

Studies have looked at mobility data from different sources for the same dataset but not different types of datasets (Huang et al., 2020b, Iacus et al., Yabe et al.). To the best of our knowledge, our research is a first of a kind that incorporates various datasets, such as population density, aggregated population mobility, intracity rail use, vehicle use, hotspot and non-hotspot movement patterns, and human activity agglomeration. The novelty of our research lies not only in performing the analysis at a finer scale but also in considering multiple datasets to reduce the uncertainties and gain multi-dimensional insights. The goal is to understand how a pandemic impacts mobility in different sectors, such as transportation—intracity rail and vehicular traffic. Looking across six different datasets, we endeavor to answer the following research questions: 1) How COVID-19 and government policies to counteract the spread of the virus influence human mobility within and across New York City boroughs? 2) How changes in overall mobility are explained by mobility through different means? 3) Does reduced mobility lead to reduced high-risk movements? The first research question gives us insights on how mobility is affected by COVID-19 awareness, state policies, and news (Figure 1) and whether these policies have a significant impact on mobility. Focusing on different datasets gives a holistic perspective on human mobility. The second research question identifies if people switch from high-risk transportation means (New York City subway) to low risk (vehicle) or show similar mobility reduction. The third research question gives insights as to whether reduced mobility causes reduced movements to and from hotspots, as movements linked to hotspots exhibit high risks of spreading the infection. Even a small minority of super-spreader POIs can account for a large number of infections (Chang et al., 2020).

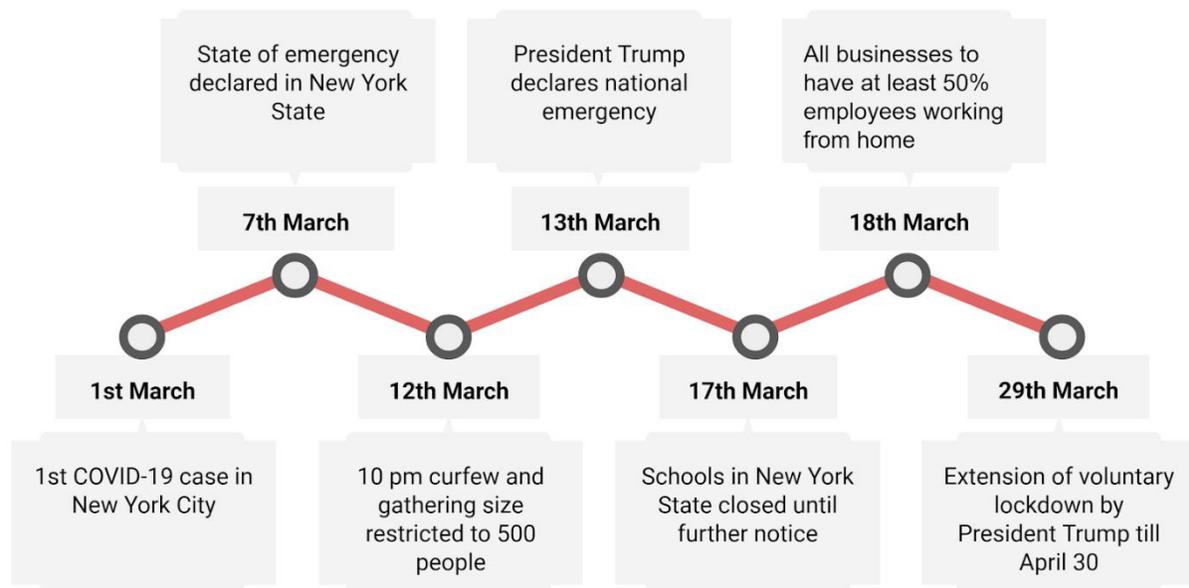

Figure 1. Timeline of policies and COVID-19-related news for the month of March for New York City and New York State

## 2. DATA AND PROCESSING

For our study, we obtained mobility data from five sources: Facebook, the Metropolitan Transportation Authority database, New York State open data portal (data.ny.gov), and from studies by Li et al. (2020) and Gao et al. (2020).

From Facebook, we acquired the Facebook Population and Movement datasets for March and April 2020. Facebook anonymizes data by adding a small random noise, implementing spatial smoothing, and dropping small counts. For these datasets, the metrics were aggregated over each tile and linked to the center of a tile or a polygon (or a combination of both). The density dataset consisted of population density, user count (number of users with the location on), among other measures for each data point. The mobility dataset consisted of start and end coordinates and the corresponding movement associated with the start-end coordinate pair. These datasets were available at a frequency of 8 hours but were aggregated over days for the movement dataset and at a weekly resolution for the density dataset. Baseline and crisis values are available for both population and movement datasets. Facebook computed the baseline values in the dataset for each data point using 5 to 13 weeks of pre-crisis data. Baseline values were calculated using data from the previous weeks of the pre-crisis period (Maas et al., 2019). We used crisis and baseline values of population density for our study. Similarly, we used crisis and baseline values for movement from the movement dataset from the start-end coordinate pairs.

We obtained MTA turnstile data from New York State open data portal for January through April 2020. The turnstile dataset consisted of the cumulative entry and exit counts for each turnstile for all subway stations in New York City at an hourly interval. To obtain the hourly entry and exit count, first the difference was computed for entry and exit time series. We then aggregated the entry and exits for these turnstiles at 1-day intervals. Tunnels and bridges toll data from MTA contained the number of transiting vehicles (cars, buses, trucks and motorcycles).

Mobility data for 16 US cities (including New York City) across points of interest classified as hotspots and non-hotspots were acquired from the study by Li et al. They used SafeGraph data to map the origin-destination network and identified the hotspots and non-hotspots on the mapped network. Here, hotspots represent high-risk zones, and non-hotspots represent the low-risk zones in terms of infection risk (Li et al., 2020). Hotspot and non-hotspot movement data for New York City were taken from this study and aggregated at the borough level for our analysis.

To consider the agglomeration of human activities, we used data Venables distance data for the study by Gao et al. (2020). They calculated Venables distance for 193 counties in the United States by using digital trace data from Mapbox. Venables distance aggregates the spatial distribution of human activities of different tiles in a county. For more information on the calculation of Venables Distance in this dataset, please refer to Gao et al. (2020). In our study, we performed the analysis at the borough level for New York City.

Table 1 gives an overview of the datasets used in this paper and unique insights offered by each of these datasets.

Table 1. Data description and insights obtained from each dataset

| Data | Information | Insights |
|---|---|---|
| Facebook density | Baseline and crisis population density values at each grid point | Changes in population density with time |
| Facebook movement | Aggregated and anonymized baseline and crisis movement between grid points | Changes in overall movement patterns across different boroughs |
| Metro turnstile | Subway entry and exit count for each turnstile for all stations in New York City | Entry and exit count for subway stations in each borough |
| MTA tunnels and bridges toll data | Number of vehicles passing through toll gates in tunnels and bridges of New York City | Count of vehicles entering and exiting different boroughs |
| Venables distance data | Agglomeration of human activities across cities | Changes in agglomeration of human activities with time for each borough |
| POI movement data (hotspots and non-hotspots) | Patterns of movement across hotspot and non-hotspot in New York City | Movement trends between hotspots and non-hotpots across different boroughs |

## 3. METHODOLOGY

This section describes methods used to analyze datasets for aspects of human mobility and to compute baseline values for Facebook Movement and subway turnstile dataset. Figure 2 illustrates the steps followed for the analysis in this paper with further explanation below.

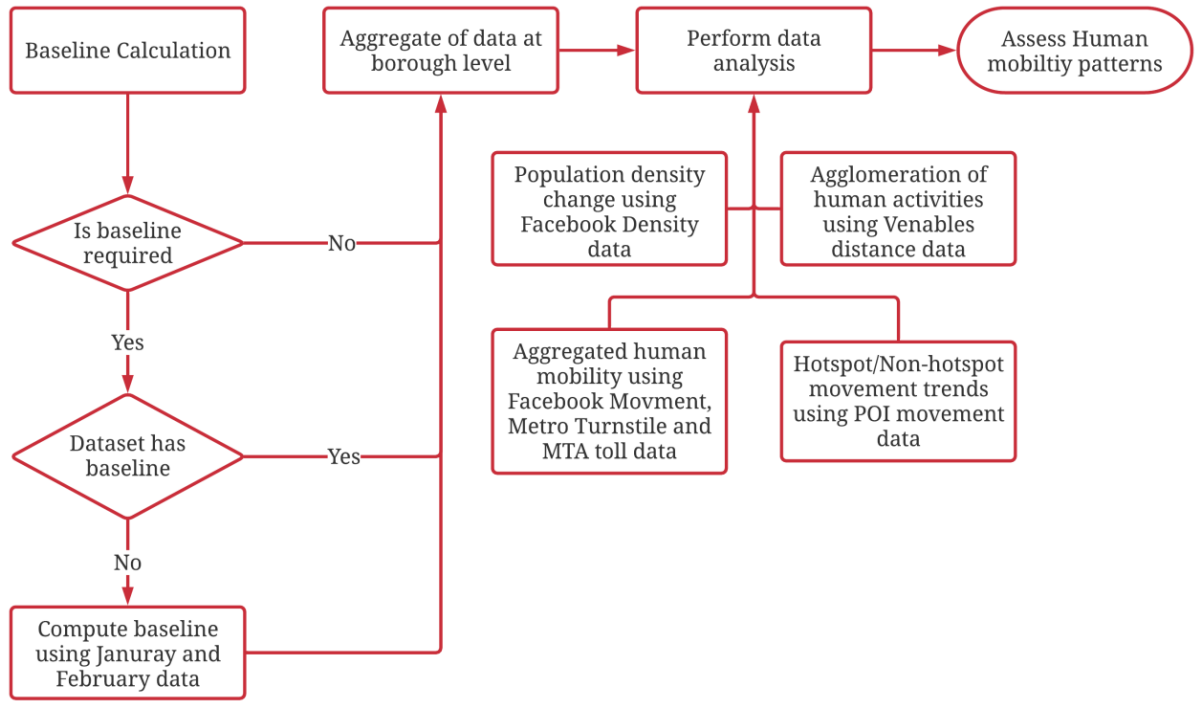

Figure 2. The process followed for analyzing human mobility using six datasets

### 3.1 Baseline calculation

Datasets, other than those acquired from Facebook, did not have baseline values from previous data. To assess the change in mobility, for Facebook Movement and Metro Turnstile datasets, the baseline values were computed by taking the average value of the data from months January and February was a combined average for both months. We used a two-month combined average instead of a day-wise average over the weeks for two reasons: 1) We did not have sufficient pre-crisis data; 2) We adopted a seven-day rolling mean approach to assess mobility for various datasets to remove the variations in mobility on weekdays and weekends, so a precise averaging was not necessary.

### 3.2 Population density

To assess the pandemic-related variation in population density for March and April, we first extracted the population density data for four main boroughs of New York City: Brooklyn, Manhattan, Queens, and Bronx. Then, we computed weekly averages of baseline and crisis values for population density from March 1through through April 30. After aggregating the data weekly for each grid point, we estimated the percentage change in population density with respect to the baseline values. We then mapped the density change for New York City for different weeks.

### 3.3 Aggregated population movement

We analyzed inter-borough and intra-borough population movement using the Facebook Movement dataset, which represents users' general mobility patterns based on the location data of mobile phones. We spatially aggregated the baseline and crisis values from Facebook Movement data at the borough level. Each start or end coordinate was assigned to a borough if it was within that borough boundary. The resampled data, therefore, instead of capturing start-end coordinate movements, captured inter-borough and intra-borough movements. Borough-to-borough movements lacking enough data points to produce statistically significant results were discarded. Then movement change time series (7-day moving average) for inter-borough and intra-borough movement for March and April were plotted.

A similar approach was adopted for the turnstile data. Analyzing subway turnstile data gives insights into how mobility in the subway compares to that observed from the Facebook Movement dataset, which corresponds to general mobility trends. MTA does not directly provide turnstile location (coordinates). We first obtained a dataset containing the coordinates of subway stations from MTA. Then we merged these with the primary dataset consisting of turnstile data for each station using station names as the common joining parameter (station names). Since these datasets were not produced in the same year, the station names had minor changes (updated name, short-form, long-form, etc.).We linked these stations manually to a coordinate. After geocoding these stations, the entry and exit data were aggregated over boroughs, applying the same approach as for the Facebook Movement dataset. The exit numbers were slightly less than the entry number for each station because people tend to use emergency exit to exit the station to save time. We therefore considered both entry and exit numbers separately for this analysis.

### 3.4 On-road movement

Restrictions on public gatherings may have a different impact on travel via personal vehicles, as they conform to the social-distancing measures and are safer from the standpoint of contact than public transport. To assess road traffic, we considered tunnel and bridge toll data as a proxy to measure the inter-borough road traffic. Since these bridges or tunnels did not connect places within a borough, we could not use this dataset to compute intra-borough vehicle statistics. All of the tunnels and bridges in New York City link different boroughs, so we grouped the toll dataset on the basis of the direction of flow and the linked boroughs. For example, incoming traffic to Brooklyn and outgoing traffic from Brooklyn were grouped separately. This gives an idea about the net inflow or outflow of traffic from a particular borough. The dataset contained the count for vehicles that pay using E-Z pass or cash (includes payment by mail). Because of the pandemic, however, some toll gates did not accept cash, and the exact date of this transition was not available publicly, so this study analyzed only total traffic flow and not the payment methods (contactless or cash). Baseline traffic values were computed for each toll plaza considering total vehicle traffic. The percent change in traffic was calculated using the baseline values for March and April. Time series plots (7-day moving average) for inbound and outbound traffic for four main boroughs were plotted for further analysis.

### 3.5 Hotspot and non-hotspot movements

Knowing movements linked to high-risk zones (hotspots) is crucial as they might contribute to the faster spreading of the infection (Almagro et al., 2020). Li et al. (2020) mapped the Origin-Destination (O.D.) networks from SafeGraph data as directed and weighted bipartite networks. They classified POIs based on a threshold in-flux and out-flux values obtained from the O.D. bi-adjacency matrix. For our study, these POI (hotspots and non-hotspots) datasets from the study by Li et al. (2020) were reclassified into different boroughs then the movements between POIs were spatially aggregated borough-wise to obtain the inter- and intra-borough movement. For each borough movement case (16 cases, – inter- and intra-borough movement), we obtained movement patterns for the hotspot to hotspot (H.H.), hotspot to non-hotspot (H.N.), non-hotspot to hotspot (N.H.), and non-hotspot to non-hotspot (N.N.) movements for March and April. This classification would give additional insights into the composition of overall movement patterns into high-risk and low-risk movements, where high-risk movements correspond to those linked with hotspots and low-risk to non-hotspot.

### 3.6 Human activity agglomeration

While higher population density and movement activity may correspond to a higher risk of infection, studying the agglomeration of human activities is also important. Higher agglomeration would imply that the average distance between people is less and could lead to a higher risk of disease spread. This measure takes into consideration the areas (ZIP code or census tract) where population density might be lower, yet still show higher agglomeration. Therefore, such areas might have a higher risk of infection spread than would be expected from assessing only population density. We obtained more-granular level data from the study by Gao et al. 2020 for New York City at the borough level to observe the activity density for Manhattan, Queens, Brooklyn, and Bronx.

**Venables distance**

Venables distance is calculated by the following equation:

$$D_v = \frac{\sum_{i \neq j} a_i(t) \times a_j(t) \times d_{ij}}{\sum_{i \neq j} a_i(t) \times a_j(t)} \qquad (1)$$

where $a_k(t)$ represents the metrics of human activities at tile $k$ at time $t$, $d_{ij}$ represents the center-to-center distance between tiles $i$ and $j$. The numerator represents the weighted distance of these activities, and, by dividing by the denominator, the value is normalized, and we obtain the weighted average, or Venables distance (Louail et al., 2014).

### 4. RESULTS

We observed the changes in population density for the four boroughs in New York City by analyzing the weekly aggregated density data from Facebook. Figure 3 shows the results for the density change (in percentage) for four boroughs of New York City—Manhattan, Brooklyn, Bronx and Queens—for March and April. We observe that in the first week of March (Figure 3 (a)) the variations in density changes are within five percent. In the second week of March (Figure 3 (b)) we begin to see a slight reduction in density in Manhattan. From the 3rd week of March (Figure 3 (c) and (d)), the changes are more apparent, and it is observed that there is a significant decrease in population density in Manhattan and other boroughs show an increase in population density. For some places in Manhattan, the reduction is more than 75%. Since Manhattan is a hub for offices and commercial spaces, results suggest that after a national emergency was declared on March 13 and office strength was reduced by 50% on March 18, people started working from home. The area of Queens showing a decrease in population density can be attributed to the location of John F. Kennedy International Airport. The airport has nearly equal passengers for both domestic and international travel (New York City data portal, 2015 data). Due to travel restrictions put on by many countries after the third week of March, people traveled less globally and also domestically, leading to lower footfall in the airport region.

It is observed that in the third week of March, the reduction in population density is the highest in comparison to previous weeks; the reduced population density in the last week of March remains similar for the month of April. The majority of the change happened in the third week of March, which coincides the announcement of a state of emergency and stay-at home orders. Moreover, the changed density state remains the same for the entire month of April (figures are available in the Appendix), which could imply that the populace effectively followed the guidelines throughout, and a steady state was achieved.

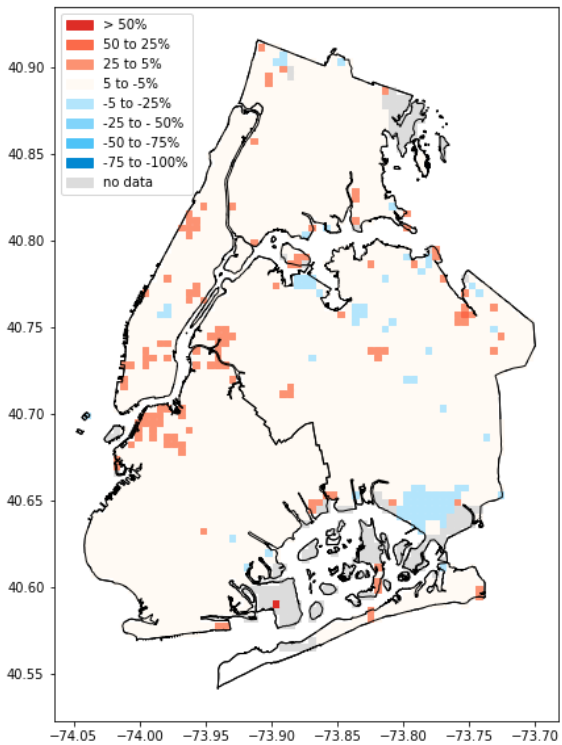
(a)

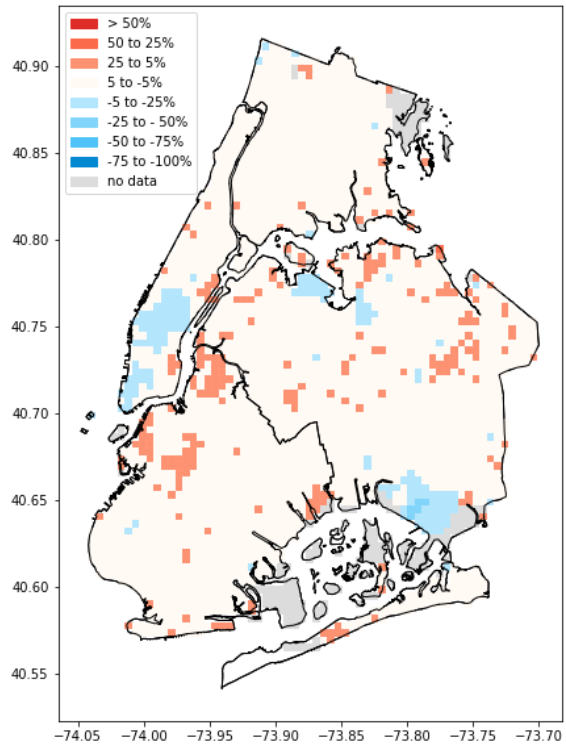
(b)

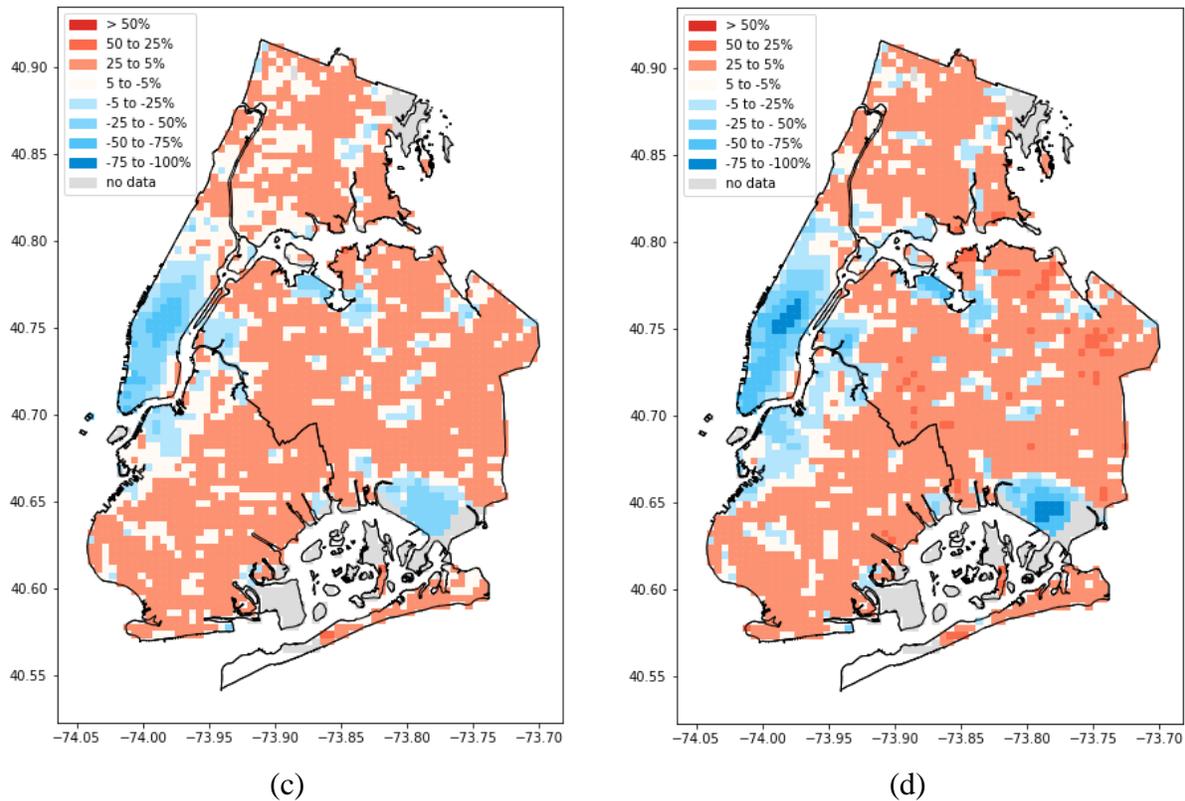

(c)                                                  (d)

Figure 3. Percent population density change (with respect to baseline values) timeline for four boroughs of New York City. (a), (b), (c), and (d) represent the population density for the first, second, third, and fourth week of March, respectively. Red represents higher population density; blue represents lower population density compared to baseline

The above results indicate a reduction in population density for Manhattan, which may reduce the risk for infection spread. For other boroughs, which show an increase in population density, it is not clear whether this increases the risk. We analyzed the mobility patterns and measured the inter- and intra-borough movement through Facebook Movement and subway turnstile data to address this question. Figure 4 shows the inter- and intra-borough movement patterns for four boroughs for New York City for general mobility using Facebook Movement data. Figure 5 shows subway ridership for each of the boroughs' stations. Figures 4 and 5 show that the reduction in mobility happens in the third and fourth week of March.

Figure 4 suggests that for the aggregated movement patterns to and from Manhattan saw a significant drop. Also, movements from Manhattan to other boroughs reduced by 60% to 80%, the largest percentage among the four boroughs. For inter- or intra-borough movements that are not Manhattan linked, the reduction was between 50% to 60%. From Figure 5, analyzing the percent change in entry counts at stations in each of the boroughs, it is evident that subway ridership dropped by more than 80% for the four boroughs. Similar to observations from Figure 4, Manhattan showed the highest drop in subway entry count (90% reduction). The use of the subway remains reduced for the month of April. This suggests that people, in general, were following the recommendations of social distancing, working from home, and avoiding the use of public transport. Movements within Bronx, Brooklyn, and Queens were slightly higher than their inter-borough movements in general, which may indicate commute for purchase of essential goods. In contrast, for Manhattan, the intra-borough movement was the lowest (compared to movement from Manhattan to other boroughs), which could be due to reduction in the population density in Manhattan; therefore, the local commute by people who come from other boroughs is comparatively less, though people working with essential services may commute to Manhattan and back from other boroughs.

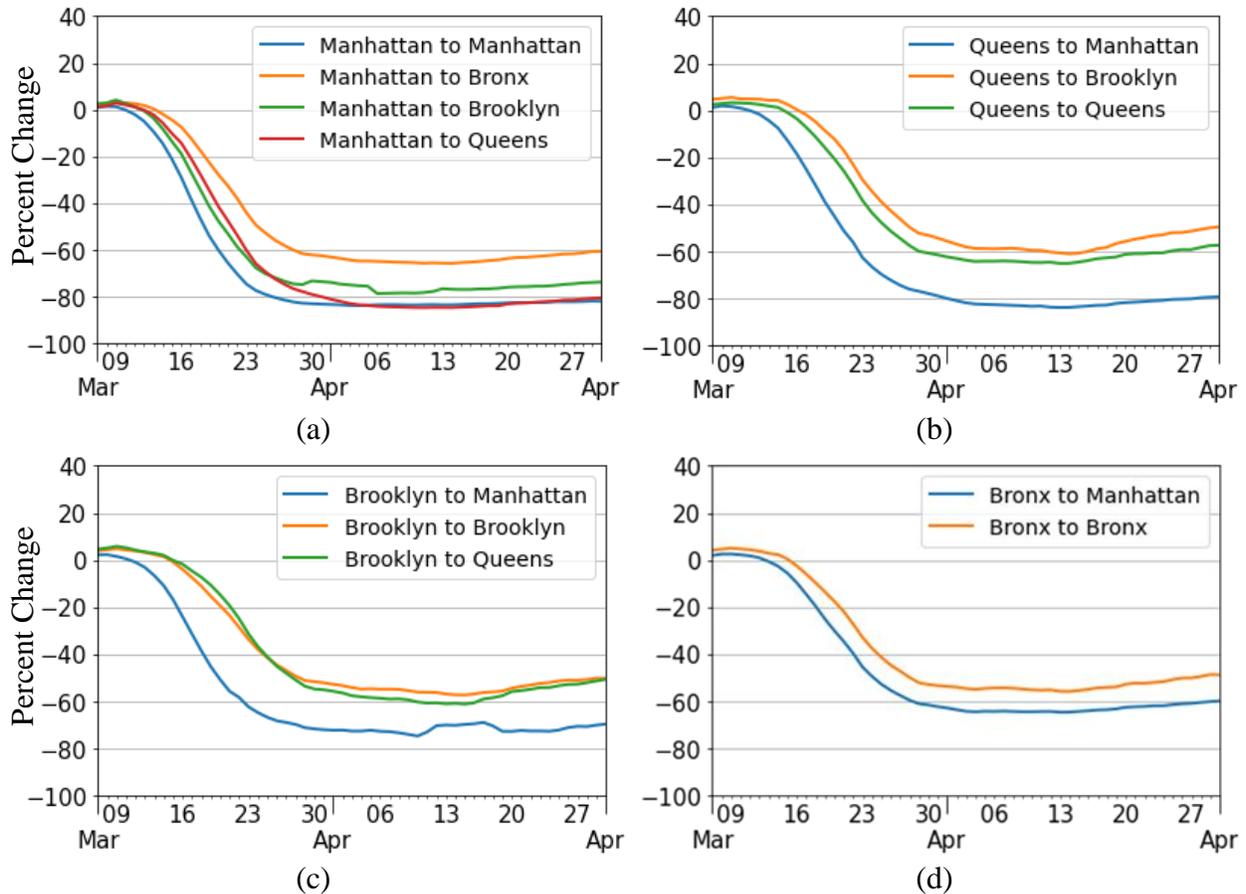

Figure 4: Seven-day rolling mean aggregated movement among four boroughs of New York City. Some borough-to-borough movement plots are not shown due to insufficient data. (a), (b), (c) and (d) show the movement from Manhattan, Queens, Brooklyn, and Bronx, respectively, to other boroughs

Results from the Facebook Movement data and Metro turnstile data indicate that while subway usage dropped by more than 80% for all the major boroughs, the overall movement within and across different boroughs showed a drop between 50% to 80%. For movements not liked with Manhattan, this reduction was only about 60%. This suggests that other means of travel might results in a lesser reduction in the overall movement than observed from the Facebook Movement data. One aspect that we have not considered yet is people traveling via road. People may resort to traveling by personal vehicles for essential shopping or for jobs exempt from having reduced workforce capacity. Analyzing mobility through roads may give additional insights into the observed difference.

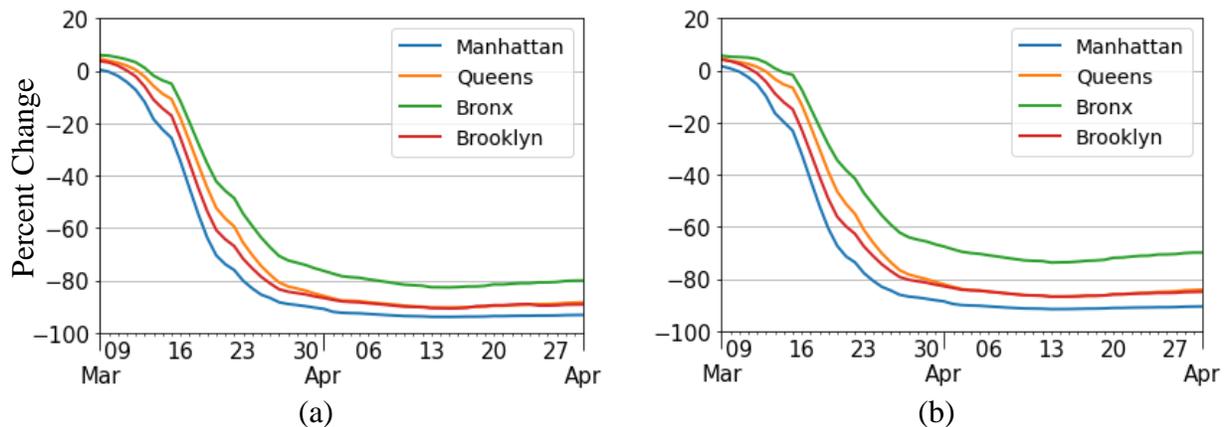

Figure 5: Seven day rolling mean of percent change of Metro ridership in four boroughs of New York City. (a) and (b) represent the percent change in entry and exit counts respectively for Manhattan, Queens, Bronx, and Brooklyn

The above results give insights into the total movement patterns and mobility through the subway system. But people involved with essential facilities may commute through roads for work, or people may travel to shop for essentials. We analyzed the tunnels and bridge toll data to observe if decreased ridership in the subway leads to an increased flow of vehicles. Figure 6 shows traffic through the tunnels and bridges connecting the boroughs. We used tunnels and bridges toll data as a proxy to study road movement. The results indicate that incoming traffic to Manhattan from the boroughs showed a drop of 70%. Incoming traffic to Queens and Bronx boroughs was reduced by about 60%. Outbound traffic to boroughs other than Manhattan shows a drop of 50% to 70%. These results show that the traffic towards Manhattan shows the highest decrease, and in general, we see about 60% less movement through roads to other boroughs.

We observe similar mobility reduction patterns in road movement as observed before for different boroughs, but the magnitude of reduction is not as much as that observed with Metro ridership. This could be because commute by car is a safer means of transportation vis-à-vis possible contagious diseases as compared to public modes of transportation. Moreover, people employed in the operation of critical facilities or emergency response would still need to commute. Additionally, this dataset accounts only for the number of vehicles that pass through a toll plaza, and there is no information about the number of passengers per vehicle. If people shared rides before the pandemic, they may not do so anymore, which could implicate slightly higher than the observed movement of cars per passenger.

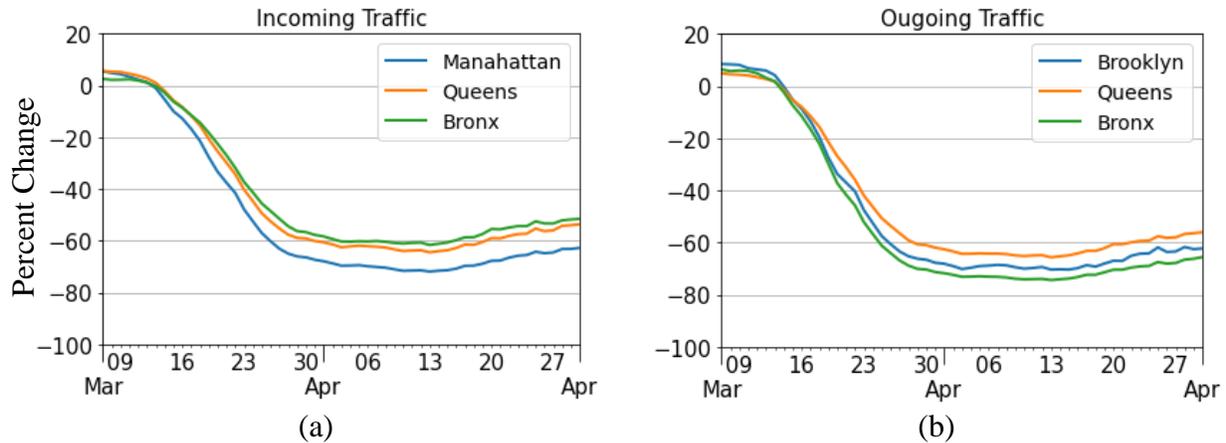

Figure 6: Seven- day moving average of (a) incoming traffic for Manhattan, Queens, and Bronx; and (b) outgoing traffic from Queens, Bronx, and Brooklyn

Our analysis so far only accounted for mobility patterns and trends for different boroughs. While it may be true that reduced mobility may mean lesser risk if there are still significant movements that are linked with hotspots, there may be some movement types that contribute to the spread of the infection. In order to determine whether reduced mobility leads to a reduction in high-risk movements, we investigated the H.H., NH, H.N., and N.N. movement patterns for the four main boroughs for New York City. Figure 7 shows the H.H., NH, H.N., and N.N. movement patterns between Manhattan, Bronx, Queens, and Brooklyn (more borough to borough movement results can be accessed in the Appendix section).

Results indicate that movement towards Manhattan involved a high number of H.H. movements. In contrast, movement from Manhattan to other boroughs involved a higher number of N.N. movements. Intra borough movements showed declining trends for H.H. movements and had low H.N. movements except for Manhattan. In other inter and intra-borough movements, N.N. movements were in higher numbers but did not show a significant drop in activity. This suggests that even though Manhattan saw the highest reduction in mobility, the risk of infection is still high as hotspot linked movements are higher in comparison to other movement pairs. Moreover, the H.N. type movement does not show a declining trend implying that people from hotspots, if infected, may infect people in non-hotpots. Other inter and intra-borough movements have dominated N.N. type of movement, which is favorable if traveling cannot be avoided. N.N. type of movement corresponds to the least risk travel, compared to other movement types in terms of likelihood of spreading the infection.

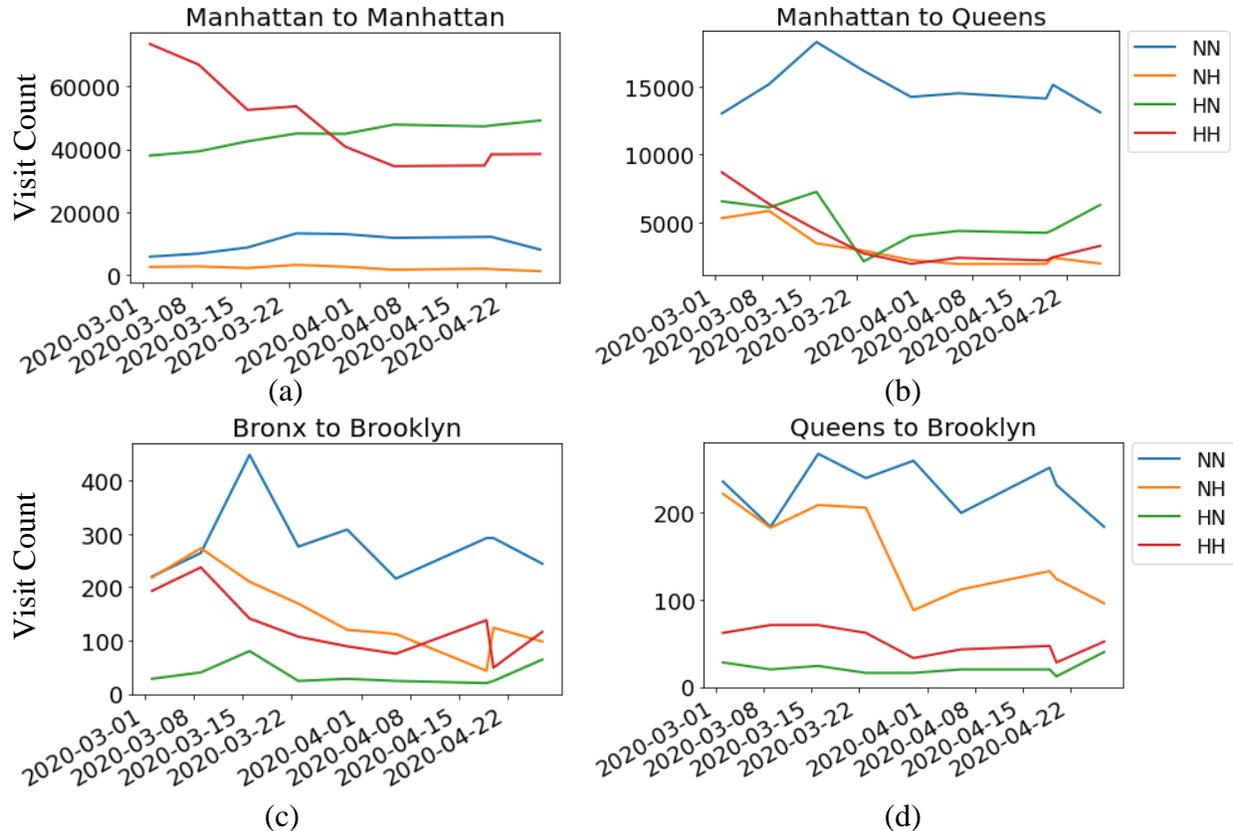

Figure 7: H.H., H.N., NH, N.N. movement patterns across four boroughs of New York City at weekly resolution. (a), (b), (c) and (d) show the movements from Manhattan to Manhattan, Manhattan to Queens, Bronx to Brooklyn, and Queens to Brooklyn, respectively. (For all figures, please refer to the appendix section 2)

To further understand the perturbed state where we have reduced mobility, we investigated the Venables distance for these boroughs to get insights into how clustered the activities are with respect to distance. Figure 8 shows the Venables Distance for four boroughs of New York City. The results suggest that the clustering of human activities increased significantly only for Manhattan in the first two weeks of March. Other boroughs showed small increment till the end of March then had a gradual decline. This indicates that Manhattan showed the most clustering of activities in comparison to other boroughs, but it was still a twenty percent increase. Combining insights from hotspot linked movement to Manhattan, results suggest that even though the distance between people is increasing, it may not reduce the risk substantially as hotspot linked movements dominate. Other boroughs like Queens, Bronx, and Brooklyn show only a marginal increase in the Venables distance.

These results indicate that the average distance between people did not change significantly for boroughs other than Manhattan. When people started working from home, they did not commute to their offices in Manhattan; hence, the Venables distance increased by roughly twenty percent only for Manhattan. For other boroughs, it is unclear why Venables distance only increases marginally. It may be because of the balancing effect of increased population density and a reduced number of trips within a borough.

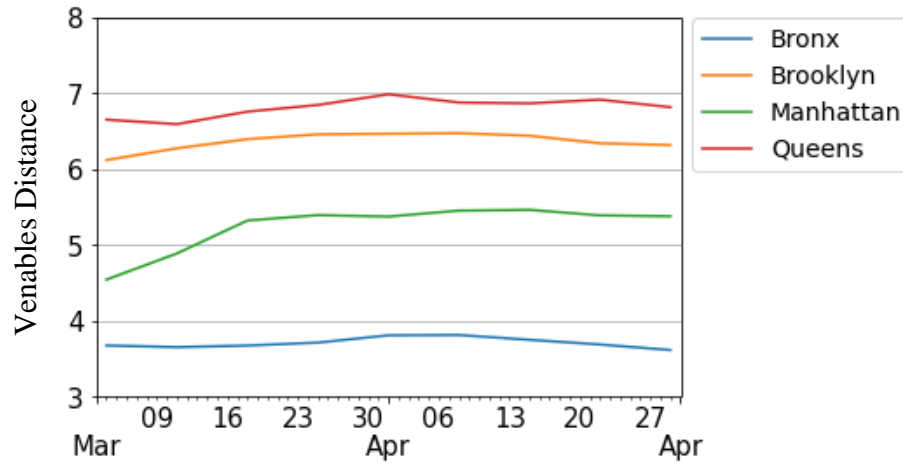

Figure 8: Venables distance time series for four boroughs of New York City

## 5. DISCUSSION

This study shows how human mobility was affected by the pandemic and how the influence exerted by the first COVID-19 positive case, government policies, and major news had on them. The estimates of aggregate flows of people can help officials understand which policies are most effective (Buckee et al., 2020). The news of the first case of the COVID-19 in New York City and the declaration of the state of emergency did not have much impact on mobility. We see a steep decrease in mobility only when the curfew is imposed on restaurants and bars by the Governor of New York followed school closures in an attempt to stop the spread of the virus. We see that in March, inter- and intra-borough movement of population declines, and towards the end of March, it reduces to a 60%-90% lower value that continues until the end of April. The highest drop in mobility is observed for Manhattan since it is a business hub and has a smaller number of residential complexes. This trend of decrease in mobility is consistent across different datasets, but we observe slight variations in the magnitude. This variation could be because data from Facebook relies on users who keep their location services enabled. While it may give valuable insights into the movement trends, we can expect some variations as they do not represent all the people. Moreover, the dataset captures general mobility trends and does not specify whether this movement is associated with traveling via road, subway, bicycle, or walking.

Subway turnstile data results indicate that the reduction in subway ridership is 10% more for Manhattan and 20% to 30% more for other boroughs. Overall, the reduction in subway ridership is more than 80% for all the boroughs, indicating the effectiveness of government control measures. Vehicular mobility shows similar patterns to that of general mobility patterns, but for boroughs other than Manhattan, the reduction is slightly more. Venables distance shows high variation only for Manhattan because it is dominated by office space and has a relatively small percentage of residential population. When the workforce started to work from home, only these areas showed clustered activities. Higher-density areas in Manhattan may be associated with residential areas. Moreover, people did not commute to work to Manhattan, which contributed to an increased average distance among people. Although we see an increase in population density for boroughs like Queens, Brooklyn, and Bronx, the mobility remains between 60% and 90%: low for both inter- and intra-borough movement. This suggests that people were commuting less frequently and strictly following the stay-at-home orders. This is substantiated by a reduction in subway ridership did not increase vehicle traffic, which showed a similar reduction in movement. But the risk of infection in Manhattan is not reduced significantly, as hotspot-linked movements still dominate even after a significant drop in overall mobility.

## 6. ACKNOWLEDGEMENT

The authors would like to acknowledge funding supports from the National Science Foundation RAPID project #2026814: "Urban Resilience to Health Emergencies: Revealing Latent Epidemic Spread Risks from Population Activity Fluctuations and Collective Sense-making. The authors would also like to acknowledge that Safegraph, Mapbox, and Facebook provided datasets under their Data for Good initiatives. Any opinions, findings, and conclusions, or recommendations expressed in this research are those of the authors and do not necessarily reflect the view of the funding agency and the data providers.